\newif\iffulledition 
\fulleditiontrue 

\documentclass[runningheads]{llncs}

\usepackage{bbding}

\usepackage[utf8]{inputenc}
\usepackage[T1,hyphens]{url}

\usepackage{graphicx}
\usepackage{subcaption}
\usepackage{amsmath}

\usepackage{tabulary}
\usepackage{tabularx}

\begin{document}

\title{From Data Disclosure to Privacy Nudges: A Privacy-aware and User-centric Personal Data Management Framework\thanks{\iffulledition This is an extended version of the following paper providing more detailed explanation to the work reported: Yang Lu, Shujun Li, Athina Ioannou and Iis Tussyadiah, ``From Data Disclosure to Privacy Nudges: A Privacy-aware and User-centric Personal Data Management Framework,'' accepted to DependSys 2019 (5th International Conference on Dependability in Sensor, Cloud, and Big Data Systems and Applications), to be held from November 12-15, 2019 in Guangzhou, China, to be published in a volume of Communications in Computer and Information Science by Springer. Please use the above citation when referring to this paper.\else An extended version of the paper can be found at \protect\url{http://www.hooklee.com/Papers/DependSys2019a_full.pdf}.\fi}}
\titlerunning{A Privacy-aware and User-centric Personal Data Management Framework}
\author{Yang Lu\inst{1}\orcidID{0000-0002-0583-2688} \and
Shujun Li\inst{1 (}\Envelope\inst{)}\orcidID{0000-0001-5628-7328}  \and Athina Ioannou\inst{2} \and Iis Tussyadiah\inst{2}}
\authorrunning{Y.~Lu et al.}

\institute{School of Computing \& Kent Interdisciplinary Research Centre in Cyber Security (KirCCS), University of Kent, United Kingdom, CT2 7NF\\
\email{\{Y.Lu,S.J.Li\}@kent.ac.uk}
\and
School of Hospitality and Tourism Management, University of Surrey, United Kingdom, GU2 7XH \\\email{\{a.ioannou,i.tussyadiah\}@surrey.ac.uk}
}

\maketitle

\setcounter{footnote}{0} 

\begin{abstract}
Although there are privacy-enhancing tools designed to protect users' online privacy, it is surprising to see a lack of \emph{user-centric} solutions allowing privacy control based on the joint assessment of privacy risks and benefits, due to data disclosure to \emph{multiple} platforms. In this paper, we propose a conceptual framework to fill the gap: aiming at the \emph{user-centric} privacy protection, we show the framework can not only assess privacy risks in using online services but also the added values earned from data disclosure. Through following a human-in-the-loop approach, it is expected the framework provides a personalized solution via preference learning, continuous privacy assessment, behavior monitoring and nudging. Finally, we describe a case study towards ``leisure travelers'' and several future areas to be studied in the ongoing project.


\end{abstract}

\section{Introduction}
\label{section:introduction}

The Internet User Stats (IWS) 2019 reported that over 56\% of the whole global population were relying on the Internet to live their lives and do business online~\cite{internetworld2019}. Being in the cyber-physical systems (CPS) where the boundary between physical and cyber worlds is disappearing, people need to disclose personal data to many external entities (organizations and other people) for using their provided services. In addition, service providers often encourage their customers to disclose more personal data for added values, e.g., special discounts or more personalized services. As a result, many people have their personal data spread over many services, and frequently become victims of data breach incidents. \iffulledition Such data breaches have been happening at a larger scale -- more frequently, impacting more users of more service providers.\fi 

The importance of protecting user privacy has also led to a widely accepted concept called ``Privacy by Design'' (PbD). The PbD concept has been officially recognized by some new data protection and privacy laws such as the EU GDPR (General Data Protection Regulation) coming into force in May 2018, which clearly defines ``data protection by design and by default'' (Article~25) as one of the explicitly listed principles. In the most developed version consisting of seven principles~\cite{privacybydesign}, two principles ``Respect for User Privacy'' and ``Visibility and Transparency'' highlight the requirement of keeping privacy user-centric, operationally transparent and visible. Existing privacy protection mainly relies on organization-facing solutions such as data leakage/loss prevention (DLP)~\cite{kaur2017comparative}. With the focus on user-centric design, privacy-enhancing technologies (PETs) are developed to address privacy issues within different applications, such as on online social networks (OSNs)~\cite{xu2017my}, cloud-computing platforms~\cite{qiu2018proactive}, mobile operating systems~\cite{mylonas2013assessing,Rastogi2015Uranine} as well as Internet-of-Thing (IoT) \cite{Das2018PPA4IoT,Tian2017SmartAuth,qiu2018proactive} environments. Despite existing methods proposed for user-centric privacy, we have observed a general lack of universal frameworks that can cover personal preference management, trade-offs between privacy risks and value enhancement\footnote{Such value refers to the added values a user can obtain by disclosing data to service providers, in addition to receiving the basic services wanted.} as well as behavioral nudging. This paper fills the gap by proposing such an ``all-in-one'' framework with the following key features:

\begin{itemize}

\item \emph{Easy bootstrapping} of the system from \emph{flexible user input} and \emph{historical data disclosure}.

\item \emph{User-centricity} achieved based on data disclosure behaviors of ``me'' (the owner and user of the framework) collected \emph{completely at the client side}, i.e., on his/her local computing device(s).

\item \emph{Being completely service-independent} as it does not introduce dependency on any external online services or a new remote service. This is important to make the solution completely user-centric and under the user's full control.

\item \emph{Trade-off analysis between privacy and added value} in the whole process, from personal preference management, joint privacy risk-value analysis, to behavioral nudging for a better trade-off between the two aspects.

\item \emph{Use of a computational ontology} to enable automatic reasoning about data and value flows, for the purposes of joint privacy risk-value analysis and nudging construction.

\item \emph{Human-in-the-loop} design enabling \emph{natural human-machine teaming} via human behavior monitoring and nudging based on technical tools.

\end{itemize}

The rest of the paper is organized as follows. The design of the proposed framework is explained in Section~\ref{section:framework-design}. Then, a case study about privacy protection of leisure travelers' data in Section~\ref{section:scenario} illustrates how the framework can be used to help a traveler to decide on disclosing personal information for added values. This can also echo the aim of ongoing project, PriVELT (\url{https://privelt.ac.uk/}) to develop a user-centric platform based on travelers' privacy-related behaviours so that effectively nudge them to share their personal data more responsibly. Finally, Section~\ref{section:conclusion-and-future-work} concludes this paper with future work.

\section{Related Work}
\label{section:relate-work}

To design a user-centric framework for data privacy protection and value enhancement, we studied related work on \emph{privacy preference learning and profiling}, \emph{privacy risk assessment} as well as \emph{privacy nudging}.

\iffulledition such work can be broadly categorized into three main areas: \emph{privacy preference learning and profiling}, \emph{privacy assessment} and \emph{behavioral nudging}.\fi

\subsection{Preference Learning and Profiling}

Privacy means differently to different people. Since Westin's Indexes segmented consumers in the \emph{Fundamentalists}, \emph{Unconcerned} and \emph{Pragmatists}~\cite{westin1991harris}, researchers have shown interest in privacy segmentation and thus developed this categorization from different aspects~\cite{sheehan2002toward,king2014taken,elueze2018privacy,park2013digital}. As the classic categorizations (Westin's Index and its variants) are questioned in predicting people’s actual behaviors, contextual factors and demographic variables were analysed and attributed to resultant clusters~\cite{elueze2018privacy,westin1991harris,peddinti2015perceived,naeini2017privacy}. Segmenting customers based on the disclosure behaviors can help system developers to understand their online customers better, customize and deliver privacy settings according to the user preference predicted. For instance, participants were requested to rank statements about privacy behaviors in technology services~\cite{lin2014modeling}. Besides, different sequences of data requests were tested to increase the prediction accuracy~\cite{woodruff2014would}. Through developing the location-sharing system ``Locate!'', participants were observed when sending real-time locations at some accuracy levels. Then, the impacts of request categories (social connections, etc.) on users' location-sharing willingness were evaluated~\cite{park2013digital}. Regarding online advertisers, Kumaraguru et al.\ concluded that participants' willingness of disclosure was affected by data sensitivity, perceived benefits as well as the relevance to advertisements~\cite{kumaraguru2005privacy}. Among the mobile users, four segments were identified based on the reported comfort levels to pairwise mobile permissions and purposes~\cite{lin2014modeling}. Similar methods were applied to study users' preferences in Internet of Things (IoT) environments~\cite{lu2018exploring} and online social network (OSN) platforms~\cite{wisniewski2017making}. In addition to profiling users with privacy preferences, we find the lack of analysing added values earned by disclosing data to service providers. Besides, insufficient work was done on adaptive preference management based on previous disclosure. As planned in the ongoing project, more details can be found in Section~\ref{section:framework-design}.


\subsection{Privacy Assessment}

Privacy impact assessment (PIA) refers to a systematic assessment which can be incorporated in decision-making in privacy management~\cite{warren2008}. \iffulledition Certain guidelines have been developed and released by organizations cross different countries~\cite{cnil2018methodology,cnil2018knowledge,cnil2018template}.\fi In a PIA template, each privacy risk can be evaluated by combining the quantities of \emph{impact} and \emph{likelihood} that it can cause~\cite{wagner2018privacy}. Specially, to assess data privacy impacts caused by data disclosure, certain processes are modelled with personal characteristics and contextual features. For instance, Alnemr et al.\ designed a data protection impact assessment (DPIA) tool based on the legal analysis of General Data Protection Regulation (GDPR) as well as the evaluation of privacy issues in cloud deployment~\cite{alnemr2015}. Noticing sensitive attributes can be collected, accumulated, and used on smart grids, Seto implemented PIA procedures for smart grids in the United States and demonstrated it could effectively visualize the privacy risks in specific activities~\cite{seto2015smart}. Towards the risks existing in publicly released datasets, Ali-Eldin et al.\ designed a scoring system containing a privacy risk indicator and privacy risk mitigation measure in data releases~\cite{ali2018opendata}. Since privacy risks can be caused by data disclosure to external entities (e.g., organizations and other users), we proposed a high-level model in Section~\ref{sec:assessment} to capture possible data flows among parties while using online services. More importantly, the proposed knowledge model highlights that the relations between data flows and added values should be considered together.

\subsection{Privacy Nudging}

Behavioral nudging refers to the use of interface elements aiming to guide user behaviors, when people are required to make judgements and decisions~\cite{weinmann2016}. Since human decision making is influenced by biases and heuristics, privacy nudging aims to help users to make better decisions without restricting their options~\cite{schneider2018}. The effects of nudging on privacy related outcomes such as willingness to disclose information or likelihood to transact with an e-commerce website have been studied in various contexts~\cite{gomez2018}. Previous studies have suggested the wide use of technical nudging interventions in order to assist users in security and privacy decisions~\cite{acquisti2017}. Also, nudging dashboards can be seen as the core in developing transparency-enhancing technologies (TETs): enable users to make decisions based on sufficient information~\cite{almuhimedi2015your,hedbom2008survey,hansen2007marrying,bier2016privacyinsight}. Therefore, any user-centric privacy protection systems should explicitly consider how such unavoidable behavioral nudges are implemented at user interface levels and try to make ethical choices for the user's benefits and with their awareness.

\iffulledition

Evidence on the effects of privacy nudging as a \emph{universal} instrument for privacy intervention, however, is still inconclusive with a number of studies demonstrating a significant effect of nudging on privacy decisions by limiting information disclosure, while others showing contradictory results. Further research is therefore essential in order to gain a comprehensive understanding on the impact of nudging on privacy outcomes. Despite the inconclusive evidence about privacy nudging in the literature, as Acquisti et al.\ pointed out in~\cite{acquisti2017}, ``whenever a designer creates a system, every design choice inescapably influences the user in some way'', so nudging is not a choice but a fact. It deserves highlighting that nudges can also be unintended or even malicious.

People are making choices every day in digital environments, such as e-government applications, buying products in e-commerce websites, booking hotels on mobile applications, etc. The interface elements embedded in all of these environments influence users’ decisions, either intentionally or sometimes even unintentionally, by how the system presents choices and organizes workflows (a practice known as choice architecture). 

\subsection{Privacy Decision and Control}

Privacy behavior research generally assumes that when making privacy decisions people make rational deliberation comparing risks and benefits associated with disclosing personal information~\cite{smith1996}. The trade-offs between benefits and risks of information disclosure are explained in the Privacy Calculus Theory~\cite{laufer1977privacy}, which has been used in various forms in privacy research, such as theories of utility maximization~\cite{awad2006personalization}, expectancy theory of motivation~\cite{stone1990privacy}, and expectancy-value theory~\cite{ajzen1991theory}. Further, in~\cite{li2012empirical} Li proposed the Risk Calculus Theory, referring to the trade-off between perceived risks and the efficacy to cope with these risks. Together, they form the Dual-Calculus model of privacy, which determines users' intention to disclose personal information. To put it simply, information disclosure is expected to be a function of a user's assessment of the risk of disclosure and the extent to which he/she could cope with such risk. Users, however, often rely on privacy seals as heuristic safety signals when transacting online, failing to notice and fully process the statements of risk~\cite{larose2007}. To that end, researchers and practitioners alike have attempted to facilitate users' comprehension of privacy risks, including ways to present risks of information disclosure in privacy warning labels. The practice of influencing users to attend closely to information regarding privacy risks can be considered privacy nudging discussed below. 

\fi

\section{The Proposed Framework}
\label{section:framework-design}

The proposed privacy-aware personal data management framework is user-centric and service-independent, designed by following the ``human-in-the-loop'' concept. By ``user-centric'' we mean that the framework has a central entity ``me'' (the user being served), whose data disclosure behaviors are monitored by technical tools. By ``service-independent'' we mean that the framework runs completely on the client side, without dependency on service providers, e.g., all processes are done in such a way that no private or sensitive data are shared with any existing or new remote service so no additional privacy issues will arise. The ``human-in-the-loop'' concept refers to the fact that in the framework the human user (``me'') provides preferences on privacy protection and add values, meanwhile a higher level of personalization is allowed via incremental and dynamic user profiling from ``historical disclosure'', and thus achieve ``user-centricity''.

\begin{figure}[!tb]
\centering
\includegraphics[width=0.9\linewidth]{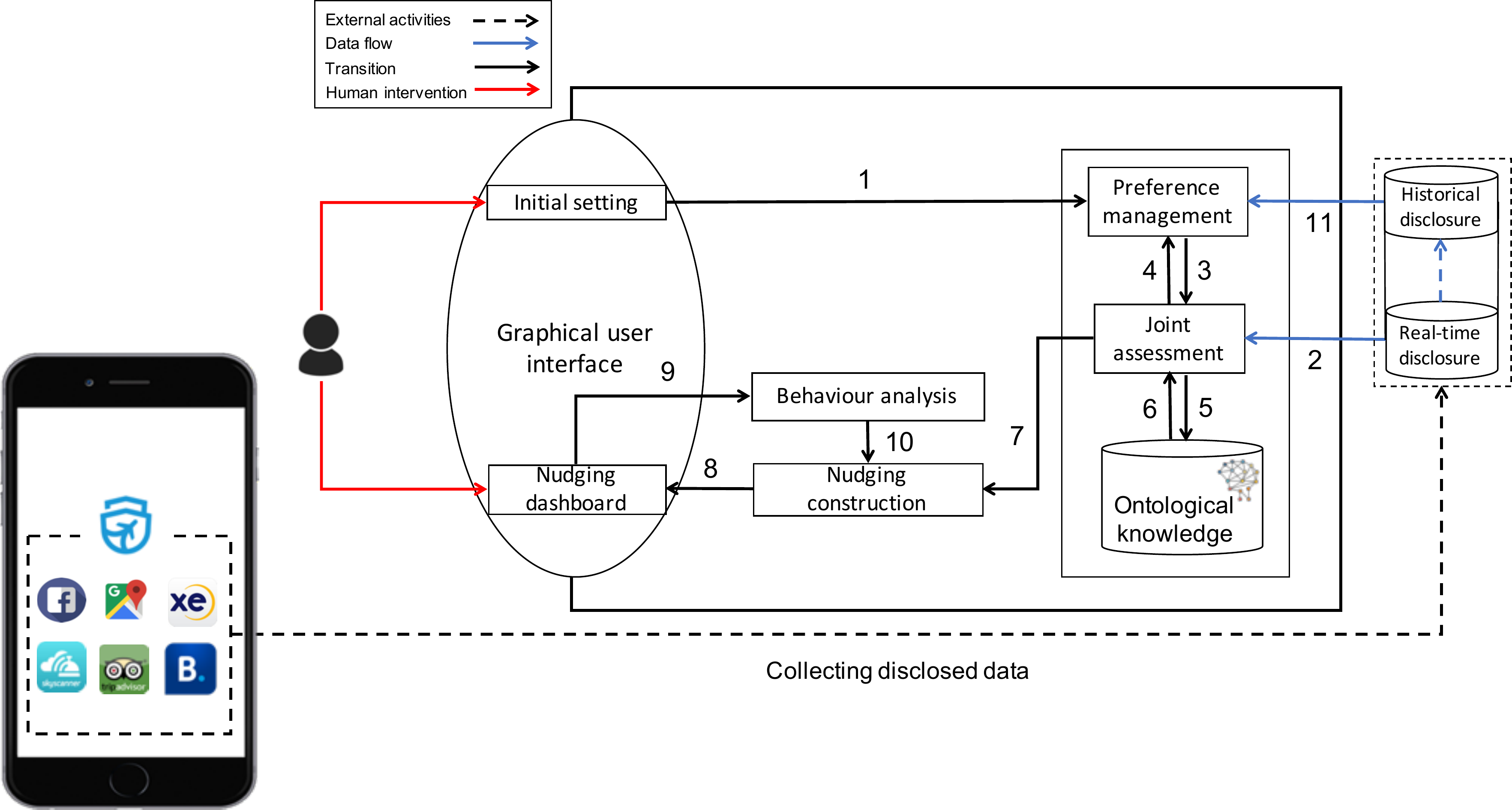}
\caption{The proposed framework with an example traveler-facing implementation}
\label{fig:high-level-design}
\end{figure}

Figure~\ref{fig:high-level-design} illustrates the high-level design of the framework with an example implementation. The overall aim is to guide the user (``me'') to manage his/her behaviors around personal data disclosure based on a better understanding of \emph{what data} have been disclosed \emph{to whom}, \emph{for what purposes}, \emph{when}, \emph{for how long}, \emph{where}, associated \emph{privacy risks} and \emph{added values} such data disclosure could bring. From the user's perspective, a typical implementation of the framework is a computer application running from the user's own computing devices, e.g., a mobile app downloaded to a smart phone for helping an individual traveler to manage data disclosure activities while travelling (a planned outcome of the PriVELT research project, shown as the blue, lock-shaped app logo in Fig.~\ref{fig:high-level-design}). Specifically, a system built on the proposed framework will have two processes running in parallel to achieve the expected functionality: 

\iffulledition

The system continuously tracks the user's behaviors and adjusts the user's profile for personalizing and contextualizing the joint privacy risk-value analysis and behavioral nudging. The system is bootstrapped with some default settings, including pre-trained AI models for various purposes (user profiling, privacy risk assessment and behavioral nudging), a data-flow knowledge base (KB) for joint privacy risk-value analysis, historical behavioral data on the user's local device and cloud servers the user has access from her/his local device, and optional input from the user on her/his privacy preferences (e.g., using a questionnaire regarding their privacy concerns). The bootstrapping process is light-weighted, and the data needed can either be simply downloaded from an online repository, collected from the user's device or a remote storage owned by the user, or directly obtained from the user.

While being service-independent, the framework can still benefit from optional collaborations of third-party service providers, e.g., retrieving more useful data from a service provider via an API into the client side for better user profiling and privacy risk assessment.

\fi

1. \emph{Operational process.} This process begins with setting the personal preference on ``privacy + value'' and the arrival of data disclosure behaviors from using external services. Based on these inputs, a joint privacy risk-value assessment component is triggered to conduct a joint privacy risk-value analysis (\emph{1--3--4}). The analysis is based on the data-flow knowledge base (\emph{5--6}), which is equipped with a computational ontology covering semantic data flows between different entities (``me'' and other entities that may consume ``my'' data and sender of ``values'' as a return of the data shared). Then, based on the user's current privacy-value preferences, real-time results about joint privacy risk and value assessment will be presented to the user (\emph{7--8}). 

2. \emph{Learning process.} By running an incremental \emph{learning process} in parallel with the \emph{operational process}, the configured settings such as initial preference and nudging templates can be presented in an adaptive manner. As shown in Fig.~\ref{fig:high-level-design}, stored data disclosure can be divided in ``Historical disclosure'' and ``Real-time disclosure'' based on a boundary pre-set. As the data gradually loses its timeliness, relatively out-dated data will be labelled as ``historical disclosure'', meaning the previous disclosure behaviors. Through monitoring \emph{which items were mostly released for added values} (e.g., more discounts in booking.com), it is possible that privacy requirements need to be lifted up or lowered down (\emph{11}). Besides, how the user interacts to nudging elements should be analyzed and in turn affect the construction (\emph{9--10}).



\subsection{Preference Learning and Management}

To a user-centric framework, personalization is the key. Therefore, the framework needs to learn about the user's privacy concerns and its privacy-value preferences. We consider the privacy risks caused by and values gained from data disclosures as conflicting aspects, and the framework aims at managing the user's preferences by providing the right trade-off. With user preferences on privacy protection and value enhancement configuring the initial environment, each user (especially laymen) can quickly set a ``baseline'' in a particular context. Specifically, standard groupings for privacy protection and value enhancement are pre-studied from sample users' input via different channels, such as online surveys, offline interviews and public data assets from online social networks (OSNs). Then, machine learning algorithms can be applied to ``categorize'' sample individuals of different profiles~\cite{zhu2015mining,lin2014modeling}. With such mappings stored in the ``preference management'' module, a new system (for a new user) can be bootstrapped by classifying the new user to an initial setting, given their ``historical data'' disclosed to other entities in cyber-physical world. Later on, the framework will dynamically adapt to the user's behavioral change, which can lead to the user being allocated to a different group or a new group created for the user if a new unique pattern is observed. For the purpose of joint privacy risk-value assessment, each such group is mapped to a profile, which can include parameters such as thresholds on acceptable trade-offs between privacy risks and value enhancement. Through comparing with the ``current preferences'', real-time nudges can be constructed for the user so he/she can learn and adapt his/her data disclosure behaviors accordingly. Such nudges can include what to do with a specific service and what service to use among a number of options.

\subsection{Joint Assessment on Privacy and Added Value}
\label{sec:assessment}

\begin{figure}
\centering
\includegraphics[width=\iffulledition\else0.65\fi\linewidth]{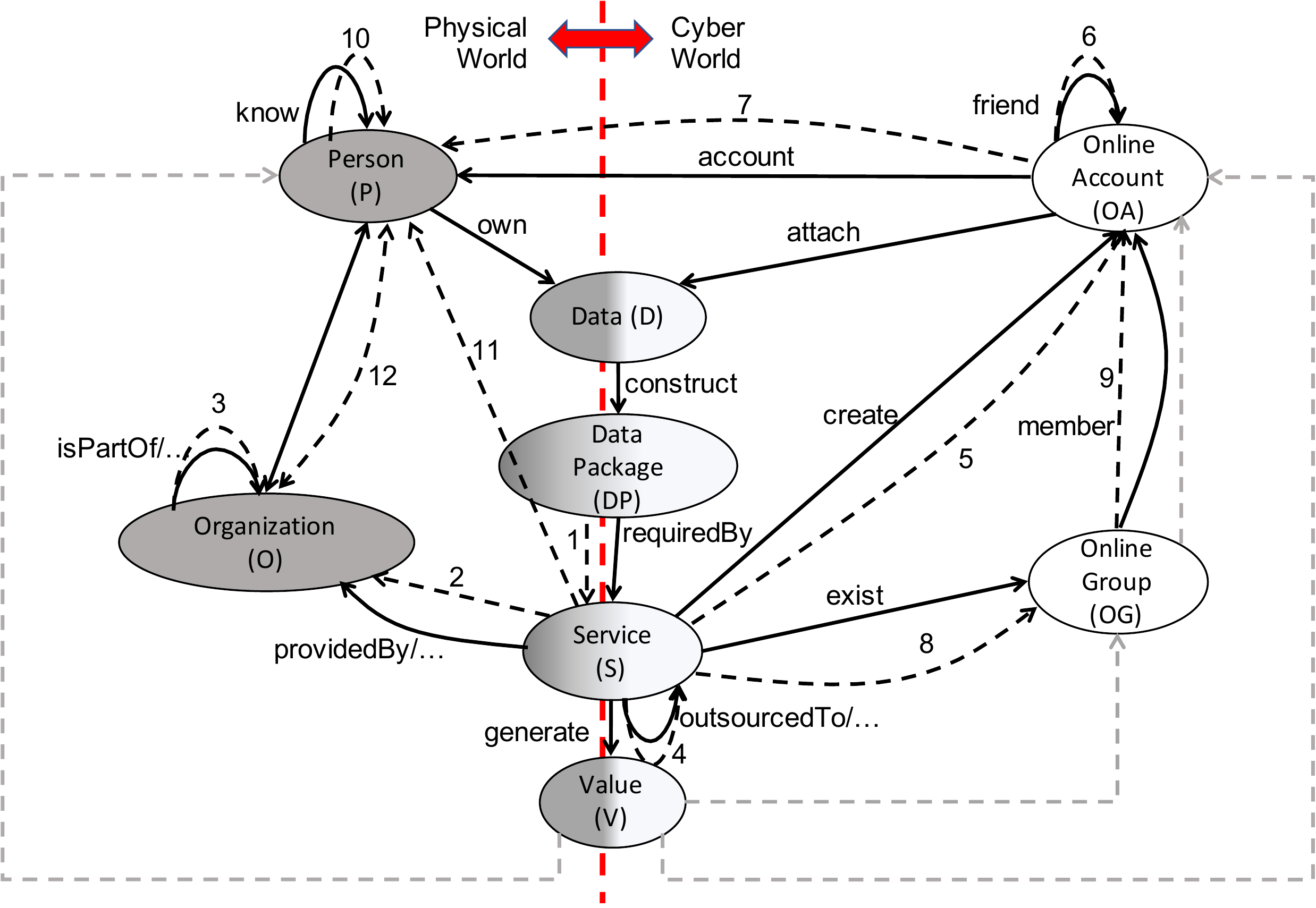}
\caption{The ontological graph of data flows in the cyber-physical world}
\label{fig:model-high-level}
\end{figure}

The joint privacy risk-value assessment process is centered around a computational ontology\footnote{The ontology described in this paper is an extended version of the one reported in~\cite{LuLi:HICSS2020}, which focuses on data flows only but not returned values.} that covers data flows on a directed graph describing how personal data of the user (``me'') can \emph{possibly} flow through (i.e., may be disclosed to) different types of entities and how the returned value (as a special type of data) can flow back to (i.e., benefit) the user in a complicated cyber-physical world. As shown in Fig.~\ref{fig:model-high-level}, the ontology includes eight essential entity types (nodes) and a number of relation types (edges) modeled in a generic, cyber-physical system (CPS) data-flow graph. Specifically, entities are categorized into three groups and colored differently: 1) physical entities (gray) that exist only in the physical world; 2) cyber entities (white) that exist only in the cyber world (from user's perspective); 3) hybrid entities (gradient) that may exist in both cyber and/or physical world. For each relation type between two entity types, there is either a semantic meaning (solid lines) or a flow (data flow: black dash line; value flow: gray dash line). Note that the graph can only show entity types and \textit{possible} relations between different entities. To analyze privacy issues, an entity level graph of entities and relations are needed to support reasoning, which will be discussed in Section~\ref{section:scenario}.
According to the graph theory, the data-flow graph can be formalized as $G=(V,E)$, where $V=\{V_1,\ldots,V_m\}$ is a set of nodes and each node $V_i$ represents an entity type treated in the same way in our model (depicted by ellipses), and $E=\{E_1,\ldots,E_n\}$ are a set of edges between nodes representing two types of relations between entities: semantic relations (represented by \textit{Type 1} edges) and data flows (represented by \textit{Type 2} edges), depicted by dashed and solid arrows, respectively. In the proposed model, there are $m=7$ different entity types (nodes) and a number of edges between them\iffulledition\footnote{These numbers can grow in enhanced versions of the model.}\fi.

\iffulledition
The different entities types currently included in the ontology include:
\begin{itemize}
\item \textbf{Person ($P$)} stands for natural persons in the physical world. The model is \textit{user-centric} as there is a special $P$ entity called ``me'' -- the user for whom the model is built. The model will include other people as well because privacy issues of ``me'' can occur due to data flows to other people who interact directly or indirectly with ``me''.

\item \textbf{Data ($D$)} refers to atomic data items about ``me'' (e.g., ``my name''). Data entities may be by nature in the physical world, or in the cyber world, or in both worlds.

\item \textbf{Service ($S$)} refers to different physical and online services that serve people for a specific purpose (e.g., a travel agent helping people to book flights).

\item \textbf{Data Package ($DP$)} refers to specific combinations of data entities required by one or more services. In this model, $DP$ entities can be seen as encapsulated data disclosed in a single transaction.

\item \textbf{Organization ($O$)} refers to organizations that relate to one or more services (e.g., service providers).

\item \textbf{Online Account ($OA$)} refers to ``virtual identities'' existing on online services. Note that even for physical services, there are often online accounts created automatically by the service providers to allow electronic processing and transmission of data, sometimes hidden from the users.

\item \textbf{Online Group ($OG$)} refers to ``virtual groups'' of online accounts that exist on a specific online service.
\end{itemize}
\fi

There are mainly two types of edges on the proposed graph. \textbf{Type 1 edges} refer to existing relations with semantic meanings that may or may not relate to personal data flows. For instance, the edge connecting the entity types $P$ and $D$ means that the special $P$ entity ``me'' owns some personal data items. Unlike Type 1 edges help model the ``evidence'' about how and why data may flow among these entities, Type 2 edges (possible data flows) can cause immediate privacy impacts. Specifically, \textbf{Type 2 edges} refer to actual data flows from a source to a destination entity. Most such edges are accompanied by a Type 1 edge because the latter constructs the reason why a data flow can possibly occur. In the following, we use $E_i$ to denote all Type 2 edges belonging to the same edge labelled by the number $i$ in Fig.~\ref{fig:model-high-level}:

\begin{itemize}
\item E1: ($DP$, $S$) flows are normally the beginning of tracking data flows in the cyber-physical system, generated by using online services.

\item E2: ($S$, $O$) flows from $S$ to $O$ entities due to the existence of Type 1 edges \textit{providedBy} in between.

\item E3: ($O$, $O$) flows between $O$ entities given the fact that one $O$entity has some relation with another, e.g., \textit{isPartOf}, \textit{invest} or \textit{collabrateWith}.

\item E4: ($S$, $S$) flows between $S$ entities due to data sharing relations between them, e.g.,  \textit{suppliedBy}, \textit{poweredBy} or \textit{outsourcedTo}.

\item E5: ($S$, $OA$) flows from $S$ to $OA$ entities due to the existence of type 1 edges \textit{create} in between.

\item E6: ($OA$, $OA$) flows between $OA$ entities given the fact that one online account is the \textit{friend} of the other.

\item E7: ($OA$, $P$) flows from $OA$ to $P$ entities due to the existence of type 1 edges \textit{account} in between.

\item E8: ($S$, $OG$) flows from $S$ to $OG$ entities due to the Type 1 edges \textit{exist} in between.

\item E9: ($OG$, $OA$) flows from $OG$ to $OA$ entities due to a specific privacy setting on OSNs, such as setting the contents are disclosed to ``group members'' only.

\item E10: ($P$, $P$) flows between $P$ entities due to the existence of type 1 edges \textit{know} in between.

\item E11: ($S$, $P$) flows from $S$ to $P$ entities directly to a person without via an $OA$ entity, e.g., a person can see public posts on Instagram.

\item E12a: ($O$, $P$) and E12b: ($P$, $O$) flows refer to data flows between $P$ and $O$ entities in both directions, each of which is due to one or more semantic relations between $P$ and $O$, e.g., a person \iffulledition owns owns or\fi works for an organization. 
\end{itemize}

\iffulledition
By analyzing the ontological graph, one can manually and automatically detect different types of privacy issues (as different topological patterns)\footnote{See Sections~3 and 4 of \cite{LuLi:HICSS2020} for more explanation on how the detection can be done.}, and semantic information about possible value enhancement that data disclosure leads to, some numeric and categorical indicators (metrics) can be derived to help prioritizing system actions (e.g., recommending to end any services causing privacy risks immediately, or holding until services cannot offer benefits to the user) and the construction of privacy nudges (discussed in greater details later in Section~\ref{subsection:nudging}).
\fi

\begin{figure}
\centering
\includegraphics[width=\iffulledition\else0.8\fi\linewidth]{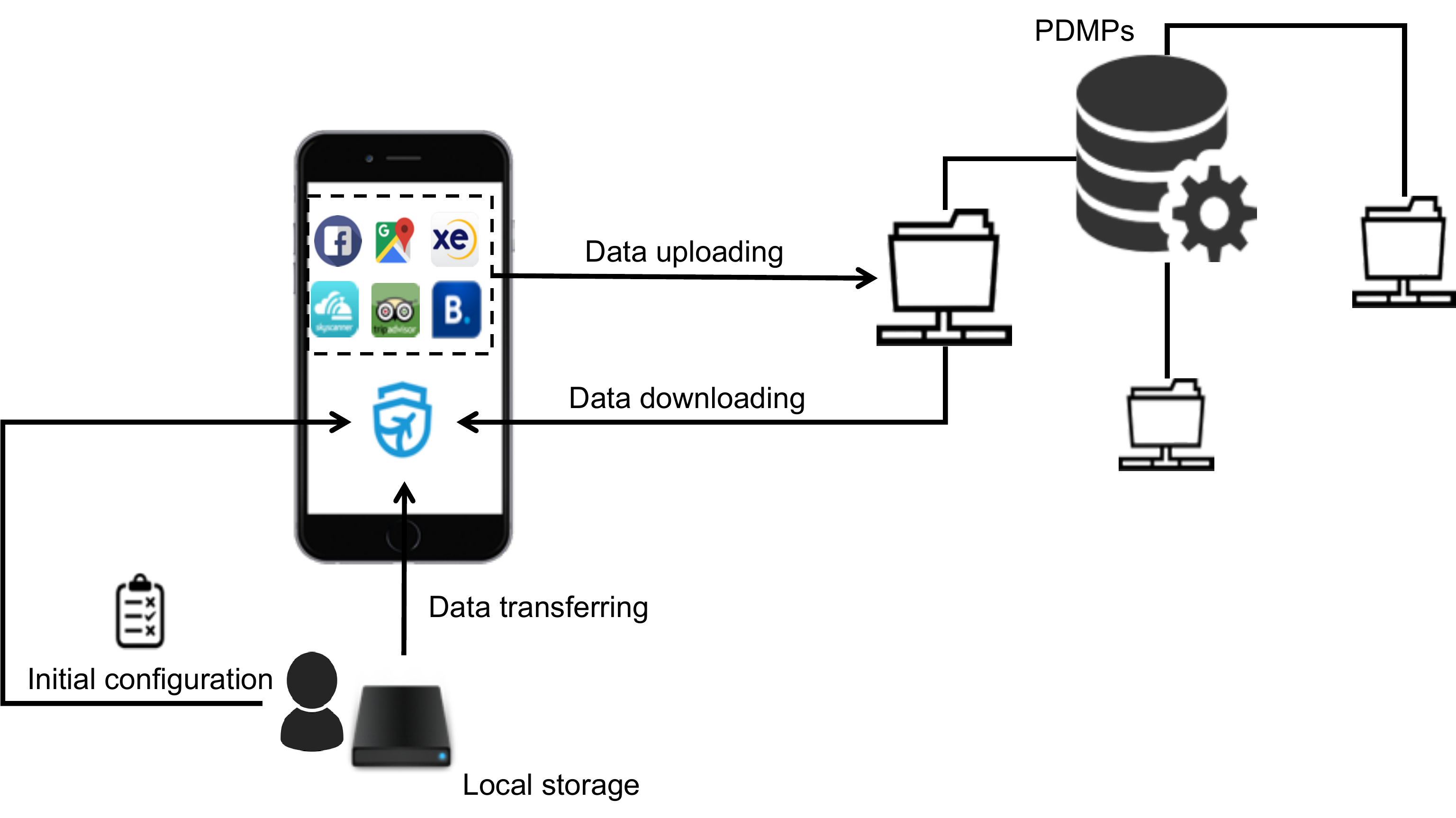}
\caption{The proposed framework working with online PDMP(s) and local storage}
\label{fig:data-collection}
\end{figure}

With the input of individual privacy preferences, (collected) data disclosure behaviors, and the entity level ontological graph about the user's data and value flows, a joint privacy risk-value assessment can be done to detect potential privacy issues, measure \iffulledition(positive and negative)\fi impacts, and recommend mitigation solutions. In order to manage data sharing with multiple entities and to reduce privacy risks, a number of user-centric personal data management platforms (PDMPs) have been developed, such as Solid (\url{https://solid.mit.edu/}), \iffulledition Databox (\url{https://www.databoxproject.uk/}), \fi Hub-of-all-things (\url{https://www.hubofallthings.com/}) and digi.me (\url{https://digi.me/}), to allow users manage their own data \iffulledition locally or in a remote data server\fi under their full control. Such platforms normally have an interface to allow new features, e.g., data analytics and visualization tools can be added so that the user can gain more insights about their data. The user of our proposed framework can decide to use one or more such PDMPs so that some (or even all) data needed for privacy risk and value enhancement assessment are stored there rather than on local devices (see Fig.~\ref{fig:data-collection}).
\iffulledition Some platforms have a particular focus on empowering the user to better manage data sharing with online services and getting values they deserve from the data shared.\fi

\subsection{Acting on Privacy Nudges}
\label{subsection:nudging}

\begin{figure}
\centering
\includegraphics[width=\iffulledition\else0.8\fi\linewidth]{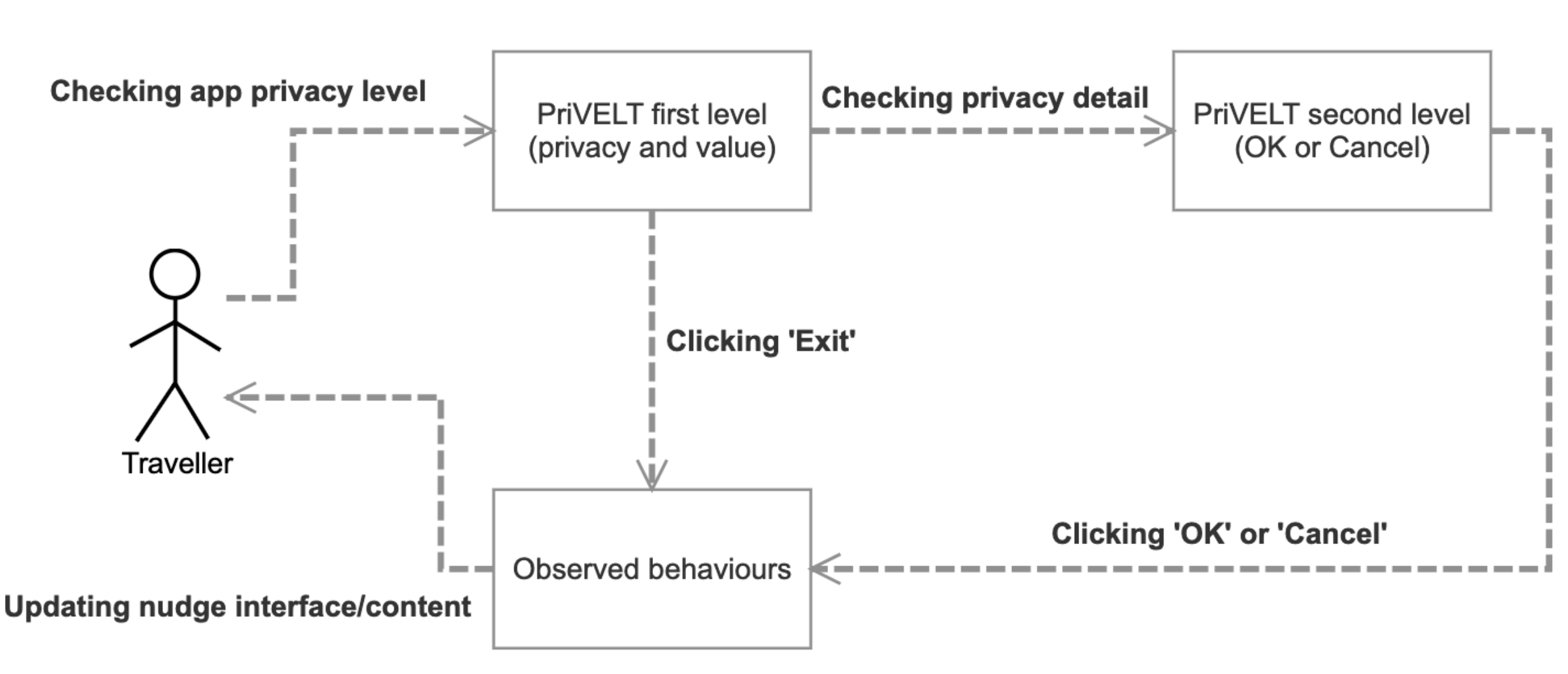}
\caption{Human interaction on two-layer nudges}
\label{fig:flow-interaction}
\end{figure}

It is widely believed that whatsoever we do with the user will have a nudging effect~\cite{acquisti2017}. For the proposed framework, rather than focusing on privacy nudging, we propose to construct privacy-value nudges, i.e., nudges that can help the user find a better trade-off between privacy risks and added values related to data disclosure decisions. In order to construct the nudges properly, it is necessary to monitor the user's actual data disclosure behaviors and his/her preferences. This task is mainly implemented through the component ``Behavior analysis''. Then, based on learnt preferences, such nudges can be constructed to deliver the expected effects, i.e. proactively avoiding risky disclosure with the knowledge about the added values to sacrifice. For instance, Figure~\ref{fig:flow-interaction} shows an example design where two-level nudging is considered: the first level is for privacy-value awareness enhancement, while mainly shows information like ``what privacy issues exist'', ``what value I have gained at what privacy costs'', ``where are the privacy issus'', and ``to what extents I should care''; the second level can be triggered to show more active interventions such as ``what options do I have'' and ``what can I do''. To the nudging contents presented on both layers, following behaviors can be monitored and analyzed to identify suitable nudging models:

1. \emph{External behaviors} refer to the behavioral change(s) after each nudge, such as switching off ``location sharing'' on the smart phone or ``delete the applications'' after being presented a nudge about an application. This is achieved from the real-time behavioral data collection part of the framework.

2. \emph{Internal behaviors} refer to the behaviors performed on the user interfaces of the proposed framework, such as counting the time-of-clicking of specific options such as ``keep it'' and ``let me know more''. This type of data is collected directly by the software implementing the framework (see one example in Fig.~\ref{fig:nudging-interface}).

\section{Case Study: Leisure Travelers}
\label{section:scenario}

Using \iffulledition two example scenarios \else one example scenario \fi about leisure travelers, we show how the proposed framework can help travelers to manage their data privacy for a better trade-off between privacy risks and enhanced experience of travel (i.e., value): \iffulledition 1)\fi booking flights and accommodations through online services\iffulledition; 2) updating travel information on social networks (OSNs)\fi. Through detecting possible data flows in using travel services, the privacy risks and enhanced travel experience by disclosing personal data can be quantified and help guide the travelers.

\begin{figure}
\centering
\includegraphics[width=\iffulledition\else0.8\fi\linewidth]{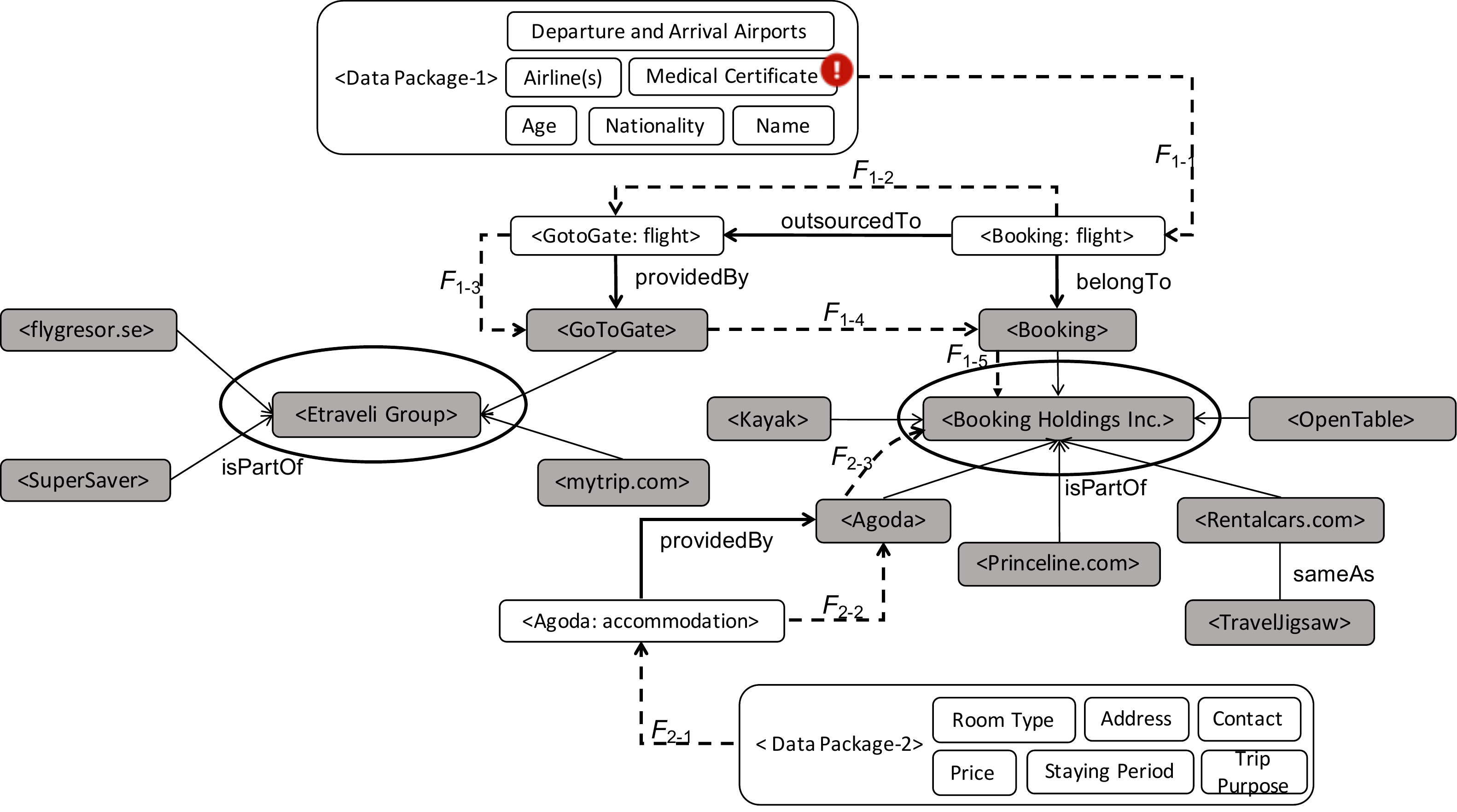}
\caption{Example ontological graph in the leisure travel context}
\label{fig:business-disclosure}
\end{figure}

In order to benefit from personalized services, personal data is often requested by travel service providers before, during, and after travel. One of the example use cases are depicted in Fig.~\ref{fig:business-disclosure} consists of entities (rectangles), relations (solid lines) and data flows (dash lines). Assuming that data submitted to use services (i.e., $F_{1\text{-}1}$ and $F_{2\text{-}1}$) are always disclosed to service providers and its parent companies\iffulledition\ (see privacy statements at \url{https://www.booking.com/content/privacy.en-gb.html})\fi, data flows $F_{2\text{-}2}$ and $F_{2\text{-}3}$ always take place so that Data Package 2 will be disclosed to the Booking Holdings Inc.\ via its subsidiary Agoda who provides the hotel booking service to the user directly. Besides, it is also important to consider special flows caused by more complex business models. For instance, the flight booking service at Booking.com is outsourced to GotoGate, which is owned by a different company group Etraveli Group. However, assuming the outsourcing contract always return the user data back to the requesting company (Booking.com in this case), data flows $F_{1\text{-}1}$, $F_{1\text{-}2}$, $F_{1\text{-}3}$, $F_{1\text{-}4}$ and $F_{1\text{-}5}$ will take place, so that Booking Holdings Inc.\ will also see Data Package 1. Now, we can see that a single company Booking Holdings Inc.\ has a more complete picture of the user's itinerary and travel preferences by combining Data Packages 1 and 2, which may not be known to the user if he/she does not know the business relationships between Agoda, GotoGate, Booking.com and their parent companies. This can create added values with privacy concerns, e.g., now Booking Holdings Inc.\ knows more about the user and can do more personalized advertising.

\iffulledition

In addition to data collection by service providers, online privacy leakage can be caused by sharing personal details with other users of the same or different services across multiple platforms (e.g., Facebook friends, Twitter followers, etc.). Sharing travel-related information across multiple platforms however could reveal data to other parties, as they can infer knowledge by joining information fragments. For instance, a traveler can upload landmark photos to Instagram while send instant itineraries on Facebook. Once these pieces are viewed by someone who owns ``friend accounts'' on both platforms, the traveler's location privacy may be threatened. Such issues can be detected by tracking data flows.

\fi

\begin{figure}[!tb]
\centering
\includegraphics[width=\iffulledition\else0.7\fi\linewidth]{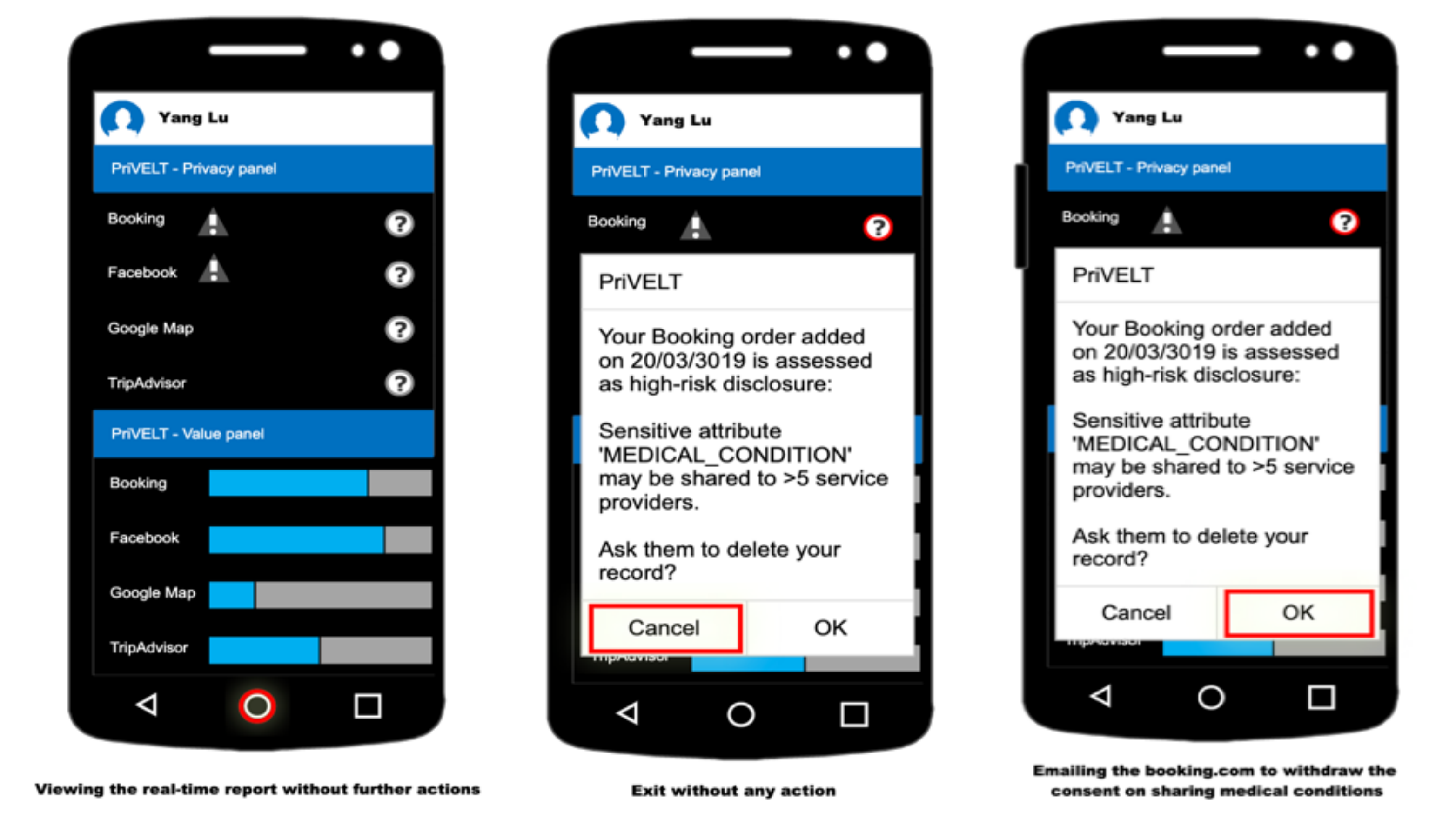}
\caption{Nudging dashboard and interfaces}
\label{fig:nudging-interface}
\end{figure}

Now let us give some example user interfaces for a different privacy issue. As shown in Fig.~\ref{fig:nudging-interface}, the first-level interface presents an overview of data disclosure activities of some ``monitored apps'', including a joint analysis of privacy risks and value enhancement those data disclosure activities lead to. For example, it shows that Booking.com is deemed a ``risky app'' but the user has also achieved a lot of benefits by using it. In addition, the example interface allows the user to ``check details without taking actions'' by clicking a question mark, which will lead the user to the second level of user interface. Given the KB and the user's disclosure behaviors, the system detects that a sensitive unit ``MEDICAL\_CERTIFICATE'' could have flowed to two different company groups and over 10 sub-companies, many of which are unknown to the user, due to special booking requirements. The system then labels this as a privacy issue after checking the user's current privacy preference. Being notified about this specific privacy issue and after considering any enhanced travel experience this may bring, the user can choose to accept the risky disclosure or request deletion of the data disclosure from some companies immediately (which may mean loss of special assistance during travel), and can adapt his/her future data disclosure behaviors accordingly. The user's choices are recorded to help personalize the user interface and future nudges, following the ``human-in-the-loop'' principle.

\section{Conclusions and Future Work}
\label{section:conclusion-and-future-work}

In this paper, we report a user-centric and privacy-aware personal data management framework, allowing a user to better manage his/her privacy in the context of interacting with multiple services and people in the cyber-physical world, via a joint privacy risk-value analysis architecture covering user preference management, joint privacy risk-value assessment, and joint privacy-value nudging. We illustrate the usefulness of the framework by using a case study about leisure travelers. Moreover, there is a number of key areas for further development of the proposed framework, which we leave as our future work.

\emph{Studying added values in different contexts\iffulledition, using travel as a highlighted context for the PriVELT project\fi.}
In which forms such ``added values'' can be represented in real-world scenarios and how they relate with data disclosure (i.e., data flows) will be studied to enrich the computational knowledge base used in the proposed framework.

\emph{Profiling travelers on their preferred balance between data disclosure and value enhancement.} To build a user-centric platform for privacy protection purposes, it is essential to learn what the privacy risks and added values mean to different users. While designing the adaptive application, this work mainly involves two facets: traveler profiling based on self-reported answers to privacy-related questions, and learning travelers' preferences from actual behaviors, which include the data disclosure and other interacting behaviors to online services and tools implementing the proposed framework.

\emph{Conceptualizing and quantifying privacy risks and added values.}
Based on data flow analysis, personal preferences and the semantic information in the knowledge base, we aim to study how to conceptualize and quantify privacy risks and added values. This will involve evolving ontological graph models and developing privacy risk indicators needed for different components such as the privacy nudge engine. Any indicators will need to cover both \emph{privacy risks} and \emph{added value}, and will need to be personalized if possible.

\iffulledition
\emph{Associating added values and privacy risks.} As discussed, the use of online services causes data disclosure, and then privacy risks, potentially. Meanwhile, users can achieve benefits from organisations (discounts, personalized services etc.) by disclosing personal information. To facilitate users' comprehension on privacy risks, the trade-off between \emph{privacy risks} and \emph{added value} needs to be interactively presented for each app.
\fi

\emph{Constructing privacy nudging based on the user's preferences.} The construction of privacy nudging should be determined by learnt personal preferences. While presenting the results from real-time disclosures, privacy nudging should give concrete and actionable recommendations such as ``which services bring more privacy risks for exchanging what added values'' and ``what can be done to mitigate such more risky services''. To effectively help privacy-related decisions, we will conduct a number of user studies to design our privacy nudging strategies and graphical elements for an implementation of the proposed framework.

\section*{\ackname}

The authors' work was supported by the research project, PRIvacy-aware personal data management and Value Enhancement for Leisure Travellers (PriVELT), funded by the EPSRC \iffulledition (Engineering and Physical Sciences Research Council)\fi in the UK, under grant number EP/R033749/1.

\bibliographystyle{splncs04}
\bibliography{main}

\end{document}